\documentclass{ws-procs9x6}
\usepackage{epsfig}
\begin{document}
\newcommand{\beq}{\begin{equation}}
\newcommand{\eeq}{\end{equation}}
\newcommand{\bea}{\begin{eqnarray}}
\newcommand{\eea}{\end{eqnarray}}
\def\simlt{\buildrel < \over {_{\sim}}}
\def\simgt{\buildrel > \over {_{\sim}}}
\newcommand{\ttbs}{\char'134}
\newcommand{\AmS}{{\protect\the\textfont2
  A\kern-.1667em\lower.5ex\hbox{M}\kern-.125emS}}

\title{Cosmological Implications of Neutrino Mass}

\author{S. F. King}

\address{Department of Physics and Astronomy, \\ 
        University of Southampton, Southampton SO17 1BJ, U.K.
E-mail: king@soton.ac.uk}


\maketitle

\abstracts{I discuss the cosmological implications of neutrino mass
including the cosmological relic density and dark matter, the 
galaxy structure limits on neutrino mass, nucleosynthesis, supernovae,
cosmic rays and gamma ray bursts. }

\section{Introduction} 
Recent SNO results \cite{Ahmad:2002jz,Ahmad:2002ka}
when combined
with other solar neutrino data especially that of Super-Kamiokande
\cite{Fukuda:2001nk}
strongly favour the large mixing angle (LMA) MSW solar solution
\cite{MSW}
with three active light neutrino states, and
$\theta_{12} \approx \pi/6$, 
$\Delta m_{21}^2\approx 5\times 10^{-5}{\rm eV}^2$
\cite{Barger:2002iv,Bandyopadhyay:2002xj,Bahcall:2002,Smirnov:2002in}.
The atmospheric neutrino data \cite{Fukuda:1998mi} is consistent with
maximal $\nu_{\mu}- \nu_{\tau}$ neutrino mixing 
$\theta_{23} \approx \pi/4$
with $|\Delta m_{32}^2|\approx 2.5\times 10^{-3}{\rm eV}^2$
and the sign of $\Delta m_{32}^2$ undetermined. 
The CHOOZ experiment limits $\theta_{13} \simlt 0.2$
over the favoured atmospheric range \cite{Apollonio:1999ae}.

The minimal neutrino sector required to account for the
atmospheric and solar neutrino oscillation data consists of
three light physical neutrinos with left-handed flavour eigenstates,
$\nu_e$, $\nu_\mu$, and $\nu_\tau$, defined to be those states
that share the same electroweak doublet as the left-handed
charged lepton mass eigenstates.
Within the framework of three--neutrino oscillations,
the neutrino flavor eigenstates $\nu_e$, $\nu_\mu$, and $\nu_\tau$ are
related to the neutrino mass eigenstates $\nu_1$, $\nu_2$, and $\nu_3$
with mass $m_1$, $m_2$, and $m_3$, respectively, by a $3\times3$ 
unitary matrix called the MNS matrix $U_{MNS}$
\cite{Maki:1962mu,Lee:1977qz}
\begin{equation}
\left(\begin{array}{c} \nu_e \\ \nu_\mu \\ \nu_\tau \end{array} \\ \right)=
\left(\begin{array}{ccc}
U_{e1} & U_{e2} & U_{e3} \\
U_{\mu1} & U_{\mu2} & U_{\mu3} \\
U_{\tau1} & U_{\tau2} & U_{\tau3} \\
\end{array}\right)
\left(\begin{array}{c} \nu_1 \\ \nu_2 \\ \nu_3 \end{array} \\ \right)
\; .
\end{equation}
Neutrino oscillation experiments give
information about the mass squared splittings  
$\Delta m_{ij}^2\equiv m_i^2-m_j^2$, and MNS parameters.
In this talk I shall show that 
cosmology and astrophysics provides complementary information
about the absolute values of neutrino masses.

\section{Cosmological relic density and dark matter}

In the early universe neutrinos were in thermal equilibrium with
photons, electrons and positrons. When the universe cooled to temperatures
of order 1 MeV the neutrinos decoupled, leading to a present day number
density similar to photons. If neutrinos have mass they contribute
to the mass density of the universe, leading to the constraint
\beq
\sum_i m_i\leq 28 \ {\rm eV}
\eeq
corresponding to 
\beq
\Omega_{\nu}h^2\leq 0.3
\eeq
in the standard notation where $\Omega_{\nu}$ is the ratio of 
the energy density in neutrinos to the critical energy density
of the universe (corresponding to a flat universe), 
and $h=H/100{\rm km}/{\rm Mpc}/{\rm s}$ is the scaled Hubble parameter.

From the current tritium limit on the electron neutrino mass 
in \cite{Bonn:rz} together with the atmospheric and solar mass splittings
we have a stronger constraint
\beq
0.05 \ {\rm eV}\leq \sum_i m_i\leq 6.6 \ {\rm eV}
\eeq
corresponding to the range 
\beq
0.0005\leq \Omega_{\nu}h^2\leq 0.07
\eeq
which implies that neutrinos cannot be regarded as the dominant component
of dark matter, but make up between 0.1\% and 14\% of the mass-energy of the
universe.

\section{Galaxy structure limits on neutrino mass}

Recent results from the 2df galaxy redshift survey
indicate that 
\beq
\sum_i m_i <1.8 \ {\rm eV} (95\%C.L.)
\eeq
under certain mild assumptions \cite{Elgaroy:2002bi}.
This bound arises from the fact that the galaxy power spectrum
is reduced on small distance scales (large wavenumber) as the 
amount of hot dark matter increases. Thus
excessive amounts of free-streaming hot dark matter would
lead to galaxies being less clumped than they appear to be on
small distance scales.

However as discussed in \cite{Hannestad:2002iz} such bounds suffer
from parameter degeneracies due to a restricted parameter space,
and a more robust bound is 
\beq
\sum_i m_i <3 \ {\rm eV} (95\%C.L.)
\eeq
Combined with the solar and atmospheric oscillation data
this brackets the heaviest neutrino mass to be
in the approximate range 0.04-1 eV. The fact that the mass of the
heaviest neutrino is known to within an order of magnitude represents
remarkable progress in neutrino physics over recent years.
This implies that neutrinos make up less than about 7\%
of the total mass-energy of the universe, which is a stronger limit than
that quoted in the previous sub-section.

\section{Nucleosynthesis}

The number of light ``neutrino species'' $N_{\nu}$ (or any light species)
affects the freeze-out temperature of weak processes which determine the
neutron to proton ratio, and successful nucleosynthesis leads to 
constraints on $N_{\nu}$ discussed in \cite{Kainulainen:2002pu}.

The limit on $N_{\nu}$ applies also to sterile neutrinos, but they
need to be produced during the time when nucleosynthesis was taking 
place, and the only way to produce them is via neutrino oscillations.
This leads to strong limits on sterile-active neutrino mixing angles
which tend to disfavour the LSND region, 
also discussed in \cite{Kainulainen:2002pu}).
Note that these limits assume that the lepton asymmetry is not
anomalously large \cite{Volkas:2000ei}.

\section{Supernovae}

It is well known that 99\% of the energy of a supernova is emitted
in the form of neutrinos, so one might hope that by studying supernovae
one can learn something about neutrino physics. Indeed studies of the
neutrino arrival times over a 10s interval from
SN1987A in the Small Magellanic Cloud have led to the 
mass bound \cite{Loredo:2001rx}
\beq
m_{\nu_e}\leq 6-20 {\rm eV}
\eeq
The LMA MSW parameter range may also have some observable
consequences for SN1987A \cite{Minakata:2001cd}.

However it should be remarked that Supernova physics
has difficulties simulating explosions, and understanding how the spinning
neutron star remnant is kick-started.

\section{Cosmic Rays}

It has been a long standing puzzle why ultra high energy (UHE) cosmic rays
are apparently observed beyond the Greisen-Zatsepin-Kuzmin (GZK) cutoff
of $5\times 10^{19}$ GeV. The problem is that such cosmic rays
appear to arrive isotropically, which would imply an extragalactic
origin, whereas neither protons nor neutrons of such energies
could have originated from more than 50 Mpc away due to their scattering
off the cosmic background photons which would excite a delta resonance
if the energy exceeds $5\times 10^{19}$ GeV.
Yet despite this, several groups have reported handfuls of events
above the GZK cut-off, and various models such as 
the Z-burst model have been invented to explain
how the GZK cut-off could be overcome \cite{Fargion:1997ft}. 
See also \cite{Gelmini:2002xy,Fodor:2001qy,Fodor:2002hy,Pas:2001nd}.
The basic idea of the Z-burst model is that UHE neutrinos
can travel much further, and scatter off the relic background
neutrinos to produce Z bosons within 50 Mpc.
Neutrino mass is required to increase the Z production rate, so
UHE cosmic rays and neutrino mass are linked in this model.

\begin{figure}[ht]
\begin{center}
\epsfxsize=5cm
\epsfysize=5cm
\epsfbox{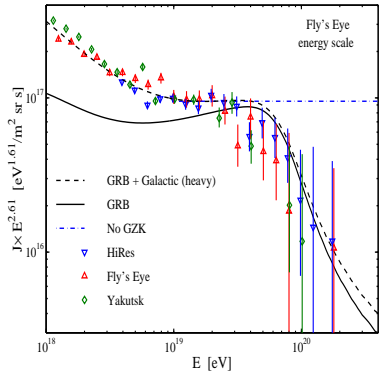}
\end{center}
\caption{\label{fig5} Recent data on cosmic rays
which shows that the GZK cut-off
might be respected after all.}   
\end{figure}

Reecently data from Fly's Eye, Yakutsk and expecially HiRes
have brought into question whether the GZK cut-off is being exceeded
at all. A recent analysis based on these data
suggests that there might be no problem with the GZK cut-off
\cite{Bahcall:2002wi}, as shown in Figure \ref{fig5}.
On the other hand Fly's Eye and HiRes establish
the cosmic ray energy from observing the fluorescence of Nitrogen
molecules using a rather dated technique.
The issue should be resolved soon by the Pierre Auger
Observatory, which can both measure the shower's pattern on
the Earth's surface, and the fluorescence of the shower in the air.


\begin{figure}[ht]
\begin{center}
\epsfxsize=10cm
\epsfysize=7cm
\epsfbox{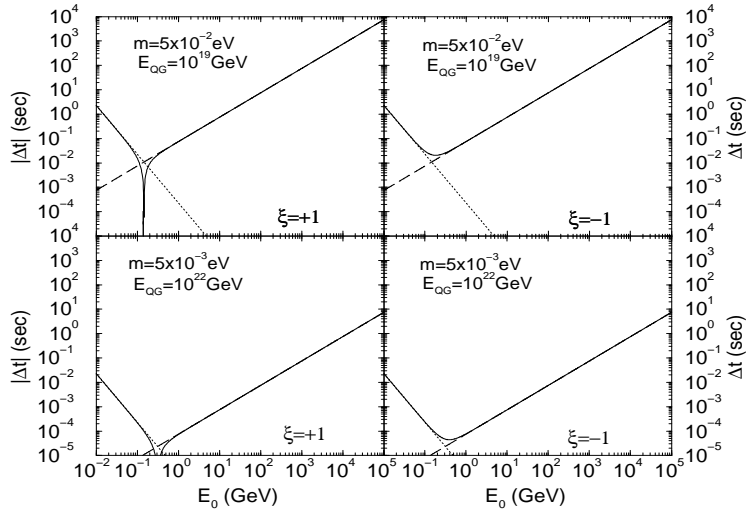}
\end{center}
\caption{\label{fig6} Time delays plotted against the neutrino energy.
The downward sloping solid curve in each panel results from 
the effect of neutrino mass dominating at low energy, while the
upward sloping solid curves are due to quantum gravity effects
dominating at high energy.}   
\end{figure}


\section{Gamma Ray Bursts (GRBs)}

GRBs are distant ($z\sim 1$), energetic and enigmatic.
According to the fireball model (for a review see \cite{Ghisellini:2001jr})
GRBs may result from the core collapse of a very massive supernova
to a compact, rotating black hole. The energy is emitted in beamed
relativistic fireball jets which are expected to contain copious
neutrino fluxes. GRBs can thus be regarded as a high intensity, high
energy neutrino beam with a cosmological baseline.

In some models of quantum gravity \cite{Ellis:1999sf} neutrinos of
mass $m$ obey the dispersion relation
\beq
p^2c^2\approx E^2(1-\xi\frac{E}{E_{QG}})-m^2c^4
\eeq
where $E_{QG}$ is the scale of quantum gravity. A consequence
of this dispersion relation is that the effects of both quantum
gravity and neutrino mass will cause time delays in the arrival
times of neutrinos compared to low energy massless photons,
as shown in Figure \ref{fig6} \cite{Choubey:2002bh}. 
Using the millisecond time
structure of GRBs it may be possible to measure neutrino masses 
from time of flight delays, in principle down to
0.001 eV \cite{Choubey:2002bh}. However this estimate assumes
complete understanding of GRBs, and large enough event rates,
and in practice there are many theoretical and experimental
challenges to be overcome.


\end{document}